
\magnification=1200
\parindent=0pt
\medskip
\centerline{\bf Quasi-rational fusion products}
\centerline{Werner Nahm}
\centerline{Bonn University}
\bigskip
abstract: Fusion is defined for arbitrary lowest weight representations
of $W$-algebras, without assuming rationality. Explicit algorithms
are given. A category of quasirational representations is defined
and shown to be stable under fusion. Conjecturally, it may coincide
with the category of representations of finite quantum dimensions.
\medskip
{\bf Introduction \hfil}
\smallskip
In conformal quantum field theories in two dimensions the role of tensor
products of Lie algebra
representations is taken over by the fusion of lowest weight
representations of $W$-algebras. This is
one of the central concepts of conformal theory, and has been much
studied from various
mathematical points of view [FF 88, KL 93, FFK 89,90, G 93].
None of the extant
approaches, however, is general enough for all purposes and applies,
e.g., to the Virasoro algebra for non-minimal values of the central
charge. Equally important, a
lot of work still has to be done to develop good fusion algorithms for
general $W$-algebras.
\smallskip
So far, the most important partial results concern rational theories on
the one hand and unitary
theories on the other. For rational theories, fusion of lowest weight
representations is
certainly calculable by existing methods, at least in principle. In
particular, the characters of
all representations can be calulcated from a finite number of terms
[MMS 88]. Once they are
known, fusion multiplicities are given by the Verlinde formula [V 88].
Apart from questions of computer memory, this is sufficient
for many purposes, though problems like
the construction of the corresponding braid group representations
require additional information.

Non-rational conformal theories are as important for physics
as the rational ones. For example, the phenomenologically interesting
Calabi-Yau theories depend on several continuous
parameters. To yield a rational theory, these must take special
values, which are not necessarily the ones required by phenomenology.
Not much is known about fusion in non-rational theories, though it may
be well behaved. Consider, e.g., the simple case of free
bosons on a torus. Here the torus metric acts as free parameter. For
generic metrics, the theories are not rational, but the fusion product
still is simple. In fact, it
just is given by the addition of the abelian charges.

Perhaps fusion always has the quasi-rational features seen for the free
bosons. It will be important
to investigate Calabi-Yau spaces from this point of view, but the tools
for this study have
yet to be developed. As a first step, we need a rigorous and convenient
definition of the fusion
product for generic theories, and better algorithms for its evaluation.
\smallskip
Unitary quantum field theories have been treated with the methods of
operator algebra [FRS 89]. Some
results reach beyond what has been established for conformal theories,
but the differences
in the axioms make the comparison somewhat difficult. In particular,
unitarity
plays no special role in the conformal context, but so far has been
crucial for the application of
operator algebras. On the other hand, conformal invariance itself is not
essential for the
latter approach.

In the following, first a concise definition of fusion will be given. If
follows the established
ideas, but puts more emphasis on generality and algorithms. A suitably
defined
category of quasi-rational representations will be shown to be stable
under fusion. These
representations have finite dimensional special subspaces, whose
dimensions behave submultiplicative
under fusion. In particular, this yields finite upper bounds on the
irreducible representations which
occur in the fusion product. Related finite upper bounds are given for
the fusion of quasirational
representations with arbitrary lowest weight representations.

The quasirational category is contained in the category of
representations with finite quantum
dimension, for which operator algebra also suggests stability under
fusion. I conjecture
that both categories are equal and sufficiently large to treat all
conformal theories. Note,
however, that theories with charges at infinity will not be considered,
since they do not fulfil all
axioms of conformal field theory. For many fusion calculations they are
essential as auxiliary
constructions, but not for the ones in this paper. In particular, the
fields of our $W$-algebras all
will have positive dimensions.
\bigskip
{\bf A definition of the fusion product}
\smallskip
For a given $W$-algebra $\cal A$ consider the
category $\cal V$ of those lowest weight modules which involve only
finitely many
irreducible representations and finitely many linearly independent
vectors of any bounded degree. By
the appropriate version of the CPT theorem, $\cal V$ admits a duality
functor which maps every
representation space $V$ to its dual $V^v$. Let ${\cal A}_0$ be the zero
mode algebra of $\cal A$ and
${\cal A}_\pm$ the subspaces of positive and negative modes of $\cal A$.
For  $V\in{\cal V}$, let
$V^l$ the subspace of $V$ which is annihilated by ${\cal A}_-$. Every
representation $V\in{\cal V}$
is determined by the action of ${\cal A}_0$ on the lowest weight spaces
$V^l$ and $(V^v)^l$. The zero
mode algebra representation on $V^l$ inherits the Lie bracket of the
$W$-algebra and an expression of
the zero modes of normal ordered products in terms of the zero modes of
the component fields. The
natural pairing between ${\cal A}_+\otimes V^l$ and ${\cal A}_+\otimes
(V^v)^l$ is the Shapovalov
form.

Most representations
encountered in conformal theory are completely reducible, which implies
$(V^v)^l=(V^l)^v$, such that
a separate consideration of $V^v$ is not necessary. In this case, $V$ is
given by ${\cal A}_+V^l$
modulo its maximal submodule which does not intersect $V^l$. More
general representations may be
treated in a similar way.

Consider some conformal theory ${\cal T}^\alpha$ with $W$-algebra
$\cal A$ and let $V_1,V_2\in{\cal V}$.
Consider a Riemann sphere with three punctures $z_1,z_2,z_3$. When
the $V_i$ are attached to
$z_i$ for $i=1,2$, they will produce some representation $V_3^\alpha$
attached to $z_3$. The fusion
product $V_{12}$ of $V_1$ and $V_2$ should yield a universal module
$V_3$ which contains all possible
$V_3^\alpha$ as factor modules and is not unnecessarily large. A
subcategory $\cal V'$ of $\cal V$  will
be called stable under fusion, if $V_1,V_2\in{\cal V'}$ implies
$V_{12}\in {\cal V'}$.

By linearity, $V_{12}$ should be a suitable completion
of $V_1(z_1)\otimes V_2(z_2)$. As a vector space, the latter space is
just
$V_1\otimes V_2$, but the action of the $W$-algebra depends on
$z_1,z_2$, which explains the notation.
To define this action, let $C$ be a small circle around $z_3$.
For holomorphic fields $\phi$ of $\cal A$, the vector $\phi(z) v$, $v\in
V_1(z_1)\otimes V_2(z_2)$
must depend holomorphically on $z$, except for possible poles at
$z_1,z_2$. Thus for 1-forms $d\omega$
which are holomorphic away from $z_1,z_2$, we must have
$$\oint_C \phi(z)d\omega (v_1(z_1)\otimes v_2(z_2))=(\oint_{C_1}\phi(z)
v_1(z_1)d\omega) \otimes
v_2(z_2) +v_1(z_1) \otimes \oint_{C_2}\phi(z) v_2(z_2) d\omega\ ,$$
where $C_1,C_2$ are small circles around $z_1,z_2$. The integrals along
them are
given by the $W$-algebra modules $V_1,V_2$.

This fact defines a natural action of the $W$-algebra on the tensor
product itself. The
commutativity and associativity properties of the tensor product will
induce corresponding
properties of the fusion product defined below. The $W$-algebra action
depends on $z_1,z_2$, but
different choices of the $z_i$ are related by rational diffeomorphisms
and lead to isomorphic modules.
When no ambiguity arises, we omit the explicit notation of the $z_i$
dependence. Using such a rational
transformation of the Riemann sphere, we put $z_3=\infty$.

A more subtle
question concerns the topology of the product space. Consider for
example vacuum
representation $V_1,V_2$ and the Virasoro field $L$. We have
$$\oint_C L(z)(z-z_1)^{-1}dz=\sum_{n=1}^\infty (z_1-z_2)^{n-1}\oint_C
L(z)(z-z_2)^{-n}dz\ .$$
Evaluating this relation on the product of two vacuum states $v_0$, we
obtain
$$(L_2v_0)\otimes v_0 =\sum_{n=2}^\infty (z_1-z_2)^{n-2} v_0\otimes
(L_nv_0)\ ,$$
which certainly is not an obvious limit in the ordinary tensor product.
Moreover, the lowest weight
vectors in $V_{12}$ will be given by similar limits.

More generally, the vectors $\oint_C \phi(z)d\omega v$,
$v\in V_1\otimes V_2$ must depend continuously on $d\omega$. For
example, we want to write
$$\oint_C \phi(z) f(z)dz v= \sum_{n=n_0}^\infty a_n \oint_C \phi(z)
z^{-n}dz v\ ,$$
where $f(z)=\sum a_n z^{-n}$ is the Laurent expansion of $f(z)$ around
$z_3$. The importance of the topology becomes obvious in the proof of
the following basic fact.
\smallskip
{\it Fusion with the basic representation leaves lowest weight
representations invariant.}
\smallskip
Proof: Let $V_1$ be the basic representation and $\phi\in V_1$. For
$v\in V_2$ we have
$$\phi(z_1)\otimes v=\oint_C \phi(z)d\omega (1\otimes v) -1\otimes
\oint_{C_2} \phi(z)d\omega v\ ,$$
where $d\omega = (2\pi i(z-z_1))^{-1}dz$. Now on $C$ we can approximate
$d\omega$ by
differentials $d\omega_N$ which are regular at $z_1$, such that
$$\oint_C \phi(z)d\omega_N (1\otimes v)=1\otimes \oint_{C_2}
\phi(z)d\omega_N v
=1\otimes \oint_C \phi(z)d\omega_N v\ .$$
Passing to the limit $d\omega$ we find
 $$\phi(z_1)\otimes v=1\otimes \phi(z_1)v\ ,$$
where $\phi(z_1)v$ is a vector in $V_2$. q.e.d.
\medskip
We circumvent the topological problems by considering the dual modules.
The
lowest weight vectors in $V_{12}^v$ are contained in $(V_1\otimes
V_2)^v$. Let $V_{12}^A$ be the
subspace of $V_1\otimes V_2$ spanned by all vectors of the form
$\phi_n v$, $n>0$, $v\in V_1\otimes V_2$. Let $V_3^l$ be the subspace of
$(V_1\otimes V_2)^v$
which annihilates $V_{12}^A$. Since the zero mode algebra ${\cal A}_0$
leaves  $V_{12}^A$
invariant, $V_3^l$ carries a representation of ${\cal A}_0$. When
$V_3^l$ is finite dimensional,
it induces a lowest weight representation $V_3\in {\cal V}$. Thus we can
define the fusion product
by  $V_{12}=V_3^v$. When $V_3^l$ is infinite, its finite dimensional
subrepresentations induce a net
of representations $V_\alpha\in {\cal V}$ which does not have a maximal
element. In this case, we
identify the fusion product with the inverse limit of the $V_\alpha$.

We now will consider a condition which guarantees that $V_3^l$ is finite
dimensional. This condition
can be checked explicitly and seems to be true for all conformal
theories for which $\cal A$ is
known. It may well be a property of all conformal theories together.
\medskip

\medskip {\bf Quasi-rational representations \hfil} \smallskip

Let $\phi$ be a $W$-algebra field of dimension $h(\phi)$. When $d\omega$
vanishes at infinity,
then $\oint_C \phi(z)d\omega$ belongs to the completion of the subspace
${\cal A}_{++}$ of $\cal A$
which is spanned by the $\phi_n$ with $n\ge h(\phi)$.  In particular, it
annihilates any lowest
weight state in the dual representation. Thus let us define $V_{12}^a$
as the span of vectors of the
form $\oint_C \phi(z)f(z)dz v$, $v\in V_1(z_1)\otimes V_2(z_2)$, where
the meromorphic function
$f(z)$ vanishes at infinity. Obviously, this space is annihilated by
$V_3^l$.

Let $V_i^a={\cal A}_{++}V_i$.  Define
special graded subspaces $V_i^s$, $V_{12}^s$ such that $V_i^s\oplus
V_i^a$ is dense in $V_i$, and
analogously for $V_{12}$. For some general considerations one might
prefer to consider the factor
space $V/V^a$ instead, but in concrete calculations one often has to
choose some special
subspace $V^s$, such that the subspace terminology is more convenient.
Notice, however, than  $dim(V^s)=dim(V/V^a)$ does not depend on any
choice.

Now we can state our first major result:
\smallskip
{\it Special subspaces are submultiplicative, in other words
$V_1^s\otimes V_2^s$ contains a special
subspace $V_{12}^s$.}
\smallskip
Proof: We prove this fact by an explicit reduction algorithm which
represents any vector
in $V_1\otimes V_2$ as a sum of vectors in $V_1^s\otimes V_2^s$ and
$V_{12}^a$.
It is sufficient to consider the dense subspace spanned by vectors
$v_1\otimes v_2$, such that the
$v_i\in V_i$ have homogeneous degrees $deg(v_i)$. The reduction
algorithm decreases the sum
$deg(v_1)+deg(v_2)$. Since the $V_i$ are lowest weight representations,
it terminates. A
special $V_{12}^s\subset V_1^s\otimes V_2^s$ then can be chosen as
complement of the intersection
with $V_{12}^a$.

We first decompose every vector into components in $V_1^a\otimes V_2$
and $V_1^s\otimes V_2$. The
space $V^a_1$ is spanned by vectors of the form $v_1=\phi_{n+h(\phi)}
v'_1$, $n\ge 0$, such that
$deg(v'_1)=deg(v_1)-n-h(\phi)$. We have  $$v_1=\oint_{C_1}
(z-z_1)^{-n-1}\phi(z) v'_1dz/2\pi i \ .$$
We replace the integral along $C_1$ by integrals along $C$ and $C_2$.
The first one gives a
contribution to $V_{12}^a$, such that we obtain the reduction formula
$$\phi_{n+h(\phi)} v'_1\otimes v_2\equiv -\oint_{C_2}
(z-z_1)^{-n-1}\phi(z) v_2dz/2\pi i
\quad modulo\quad V_{12}^a\ .$$
Since the differential is holomorphic inside $C_2$, the right hand side
is easy to evaluate and yields
$$v_1\otimes v_2\equiv v'_1\otimes\sum_{k=1}^\infty (-)^n{n+k\choose
k-1}(z_1-z_2)^{-n-k}
\phi_{h(\phi)-k}v_2\ .$$
Thus it is a sum of vectors of degrees $deg(v_2)+h(\phi)-k+deg(v'_1)$,
all less than
$deg(v_1)+deg(v_2)$. Vectors in  $V_1^s\otimes V_2$ are decomposed into
components in
$V_1^s\otimes V_2^s$ and $V_1^s\otimes V_2^a$. The latter are reduced
analogously. q.e.d.
\smallskip
In particular,
$$dim(V_{12}^s)\le dim(V_1^s)dim(V_2^s)\ ,$$
such that the subcategory ${\cal V}^q$ given by the modules $V\in{\cal
V}$ with
finite $dim(V^s)$ is stable under fusion. The modules of this
subcategory will be called
quasirational. If ${\cal V}^q={\cal V}$, then $\cal A$ itself will be
called quasirational.
The vacuum representation $V_0$ always belongs to ${\cal V}^q$, since
$V_0^s$ is just the one
dimensional ground state, in agreement with the result that $V_0$ is the
identity of the fusion
product.

So far, no conformal theories without quasirational $W$-algebra has been
shown to exist, though for
generic Calabi-Yau theories the question remains open.
\medskip
If some $V$ belongs to ${\cal V}^q$, it is possible to give a
constructive proof of this
fact.  Consider a complete set of simple fields $\phi^i$ which generates
all others by linear
combinations, derivatives, and normal ordered products.
 \smallskip
{\it Proposition: Let $V'$ be invariant modulo $V^a$ under the action of
the Fourier components
$\phi^i_n$, $n<h(\phi^i)$. Then $V'$ contains a special subspace.}
\smallskip
We have to check that $V'+V^a$ is dense in $V$. Since $V'$ generates
$V$, the latter space is
spanned by vectors $v\in\cal V$ which are given by the action of
monomials in field components
$\phi^i_{n}$ on graded vectors $v'\in V'$. Any such vector will be
reduced to a sum of vectors in
$V'$ and $V^a$ by the following algorithm. Commute all field components
in ${\cal A}_{++}$ to the
left and collect vectors in $V^a$. For the remaining monomials, act with
the rightmost field
component on $v'$ and rewrite the result as a sum of vectors in $V'$ and
$V^a$, which is possible by
assumption. Then iterate the procedure. To show that this algorithm
terminates, let the complexity of
such a vector $v$  be given by the sum of the conformal dimensions
$h(\phi^i)$ of all monomial
factors plus $deg(v')$. Taking commutators lowers complexity, such that
the first reduction step
works. Subsequent acting with $\phi^i_n$ on $v'$ yields a linear
combination of a vector $v''\in V'$,
and vectors of the form $\phi^j_mw$. The former has degree
$n+deg(v)<h(\phi_i)+deg(v')$, the latter
have degrees $h(\phi^j)+deg(w)\le n+deg(v)$. Thus in each step the
complexity decreases, and the
algorithm terminates after reducing $v$ to a finite sum in $V'+V^a$.

Conversely, if $V'$ contains a special subspace, then $V'+V^a$ is dense
in $V$, such that $V'$
fulfils the condition of the proposition. Now for sufficiently large $N$
any finite dimensional $V^s$
is contained in the subspace $V^N$  spanned by the vectors of degree
less than $N$. When the
representation is quasirational, then $V^N$ will satisfy the condition
of the proposition for
sufficiently large $N$. Since these spaces are finite dimensional,
quasirational representations can
be recognized by finite computations, whenever there is a finite
complete set of simple fields
$\phi^i$. The $W$-algebras of conformal theories all seem to have this
property.

On the
other hand, one still has to find general procedures to prove that some
representation is not
quasirational. Explicit expressions for the Shapovalov form are
certainly sufficient to do this, but
in general cases they may be difficult to obtain.

The intersection $V_{12}^\sigma$ of $V_1^s\otimes V_2^s$ with $V_{12}^a$
will be called the spurious
subspace. The action of the $W$-algebra gives an injective map from
${\cal A}_{++}\otimes V_{12}^s$ to
$({\cal A}_{++}\otimes V_1^s)\otimes ({\cal A}_{++}\otimes V_2^s)$. When
the natural maps from
${\cal A}_{++}\otimes V_i^s$ to $V_i/V_i^s$ are injective, in another
terminology when there are no
null field relations in the $V_i$, then the spurious subspace vanishes
and the calculation of the fusion products is straightforward.

More generally, consider the spaces of null field relations as modules
of the algebra
${\cal A}^{++}$ generated by ${\cal A}_{++}$ and find bases (more
precisely, generating sets, since
the module is not free). When $R_1,R_2$ are such relation bases  for the
representations $V_1,V_2$,
one has to reduce $R_1\otimes V_2^s$ and $V_1^s\otimes R_2$ to a vector
space of relations in
$(V_1^s\otimes V_2^s)+ V_{12}^a$ according to the algorithm described
above. That relation space
produces $V_{12}^\sigma$.

Relations are given by the kernel of the Shapovalov form, but without
knowledge of this form
it may be difficult to prove that a generating set of relations is
complete. Nevertheless,
for a given set of relations the previous calculations yield lower
bounds on $V_{12}^\sigma$. On the
other hand, braid group representation theory yields upper bounds. When
the bounds coincide, the
spurious subspace is known and the fusion product is completely
determined by $V_{12}^s=(V_1^s\otimes
V_2^s)/V_{12}^\sigma$. In concrete cases, this procedure seems to work
quite well, but it still is an
open problem to devise an algorithm which always will do the job.
\medskip
Operator algebra results suggest that the category ${\cal V}^f$ of
representations
with finite quantum dimensions also is stable under fusion, since
quantum dimensions behave
multiplicatively under fusion. Since they are $\ge 1$ and additive under
direct sums,
all representations in  ${\cal V}^f$ involve only a finite number of
irreducible ones, such that
one should have ${\cal V}^f\subset{\cal V}$. To make general use of
operator algebra arguments,
however, one should remove the unitary constraint and prove that the
quantum dimensions of conformal
theory coincide with those of the operator algebraic approach.

It also would be nice to prove ${\cal V}^f={\cal V}^q$. I don't know any
counterexample, but only
${\cal V}^q\subset {\cal V}^f$ is obvious on first sight. Indeed,
$V^s+{\cal A}_{++}V^s$ is
dense in $V$. On the other hand, the vacuum representation is
essentially given by
${\cal A}_{++}$ itself. Thus the quantum dimension of the $V$ is $\le
dim(V_i^s)$. Equality holds,
when there are no null field relations in the $V_i$. In the latter case,
the multiplicative behaviour
of the quantum dimensions is obvious.

Let us consider explicit calculations of the fusion product. For
simplicity, we only
consider irreducible representations. Let
$V_2^A$ be the space of vectors of the form $\phi_n v$, $n>0$, such that
$V_2^l\oplus V_2^A$ is dense in $V_2$.
\smallskip
With a slight complication of the previous arguments we prove
$$V_1\otimes V_2 \equiv V_1^s\otimes V_2^l\quad  modulo\quad V_{12}^A\
,$$
using an explicit reduction algorithm from ${\cal A}_{++}V_1^s\otimes
{\cal A}_+V_2^l$ to $V_1^s\otimes V_2^l$.
\smallskip
Proof: Let us put $z_2=0$. For homogeneous vector $v_1\otimes v_2$ we
reduce $deg(v_2)$,
or $deg(v_1)$ for fixed $deg(v_2)$.
We first reduce to $V_1^s\otimes V_2$. Thus we consider
$v_1=\phi_{h+n}v'_1$, $h=h(\phi)$, $n\ge 0$.
We have
$$\eqalign{&v_1\otimes v_2=\oint_{C_1}(z-z_1)^{-n-1}\phi(z)dz v'_1/2\pi
i\otimes v_2 \cr
&\equiv v'_1\otimes \oint_{C_2}(z-z_1)^{-n-1}\phi(z)dz v_2/2\pi i \
.\cr}$$
Now we expand $(z-z_1)^{-n-1}$ in a Taylor series around 0. This yields
$$\eqalign{&v_1\otimes v_2\equiv\cr
&(-)^{n+1}\sum_{k=0}^{h-2}{n+k\choose n}z_1^{-n-k-1}\oint_{C_1} z^k
\phi(z)dz
v'_1/2\pi i\otimes v_2\cr
&+(-)^n v'_1\otimes \sum_{k=h-1}^\infty{n+k\choose
n}z_1^{-n-k-1}\oint_{C_2} z^k \phi(z)dz
v_2/2\pi i \ .\cr}$$
In the first term on the r.h.s. we expand $z^k$ as a polynomial in
$z-z_1$.
Altogether this yields
$$\eqalign{&v_1\otimes v_2\equiv\cr
&(-)^{n+1}\sum_{k=0}^{h-2}\sum_{l=0}^{h-2-k}{(n+k+l)!\over
n!k!l!}z_1^{-n-k-1}
\phi_{h-k-1} v'_1\otimes v_2\cr
&+(-)^{n+1} v'_1\otimes\sum_{k=h-1}^\infty{n+k\choose
n}z_1^{-n-k-1}\phi_{h-k-1} v_2\ .\cr}$$
The terms on the right hand side of this reduction formula have either
lower degree in $V_2$ or fixed
degree in $V_2$ and lower degree in $V_1$. Iterating the procedure we
end up with
$v_1\in V_1^s$, and we only have to consider $v_1\otimes v_2$ with
$v_2=\phi_n v'_2$, $n>0$.
As above
$$v_1\otimes v_2\equiv \sum_{k=1}^\infty (-)^k{n-h+k\choose
k-1}z_1^{-n+h-k}
\phi_{h-k}v_1\otimes v'_2\ .$$
Since $v'_2$ has lower degree than $v_2$, the algorithm will terminate.
q.e.d.
\smallskip
 As before,
one obtains the corollary
$$dim(V_{12}^l)\le dim(V_1^s)dim(V_2^l)\ .$$
In particular, this implies that $V_1\in {\cal V}^q$, $V_2\in {\cal V}$
implies $V_{12}\in {\cal V}$.
In other words, the category of general lowest weight representations
remains stable under fusion
with quasi\-rational representations.
\smallskip
The formula
$$\phi_0(v_1(z)\otimes v_2)=\sum_{k=0}^{h-1}{h-1\choose
k}z^k(\phi_kv_1)(z) \otimes v_2 +
v_1(z)\otimes \phi_0v_2$$
yields the explicit representation of ${\cal A}_0$ on  $V_1^s\otimes
V_2^l$  modulo its intersection
with the spurious subspace, once the right hand sight is reduced as in
the previous calculation.
More precisely, one finds an explicit map from ${\cal A}_0\otimes
V_1^s\otimes V_2^l$
to $V_1^s\otimes V_2^l$ itself, once
the projections of the vectors $\phi_n v$, $n<h(\phi)$, $v\in V_1^s$ to
$V_1^s$ and $V_2^a$ have been
calculated. All of this is given by finite computations, but due to the
possibility of spurious
vectors only gives upper bounds on the fusion multiplicities.

Some spurious representations can be thrown out quickly. If $R$ is a
normal
ordering relations in ${\cal A}_0$, then the image of $R\otimes
V_1^s\otimes V_2^l$ under the
reduction procedure is spurious.
Moreover, the commutativity of the fusion product allows to remove all
irreducible subrepresentations
of ${\cal A}_0$ on  $V_1^s\otimes V_2^l$ which do not occur in
$V_1^l\otimes V_2^s$. Still, the
example of the Kac-Moody algebras shows that this is not sufficient. For
them, $V^s=V^l$, such that
our explicit algorithm just reduces to the calculation of the ordinary
tensor product of the
underlying finite dimensional Lie algebras and does not yet detect the
spurious subspace,
which requires explicit consideration of null field relations.
\bigskip
{\bf Examples}
\smallskip
Let us  illustrate the algorithm for some representations of the
Virasoro algebra with
$c=1-6(p-q)^2/pq$. Let $h(r,s)=((pr-qs)^2-(p-q)^2)/4pq$, wit $p,q,r,s$
not necessarily integral.
Let $V_{r,s}$ be the irreducible module with lowest weight $h(r,s)$. For
integral $r,s$ the
representations are minimal or degenerate and in any case quasirational,
whereas the other
representations have infinite quantum dimensions.

In $V_{2,1}$ the ground state $v_1$ satisfies
$$L_1^2v_1={2(2h(2,1)+1)\over 3}L_2v_1\ .$$ The space $V_{2,1}^s$ is
spanned by $v_1$ and
$L_1v_1$, since $L_1^2v_1$ already belongs to $V_1^a$. In particular,
the representation $V_{2,1}$
is quasirational, even if it is just degenerate and not minimal.
Consider the fusion product with some $V_{r,s}$ with lowest weight state
$v_2$. Since ${\cal A}_0$
reduces to $L_0$, we just have to calculate
 $$L_0(v_1\otimes v_2)=(h(2,1)+h(r,s))(v_1\otimes v_2)+z_1(L_1v_1\otimes
v_2)$$
and
$$L_0(L_1v_1\otimes v_1)=(h(2,1)+h(r,s)+1)(L_1v_1\otimes
v_2)+z_1(L_1^2v_1\otimes v_2)\ .$$
The right hand side of the latter expression reduces to
$$L_0(L_1v_1\otimes v_2)\equiv{2h(r,s)(2h(2,1)+1)\over
3}z_1^{-1}(v_1\otimes v_2)+
{3h(r,s)-h(2,1)+1\over 3}(L_1v_1\otimes v_2)\ .$$
Diagonalization immediately yields the eigenvalues $h(r+1,s)$ and
$h(r-1,s)$ of $L_0$ on
$V_{2,1}^s\otimes V_{r,s}$, as expected. For minimal representations,
one of the eigenvalues may be
spurious. For degenerate representations, the quantum dimensions are
integral and immediately
calculable. For generic $c$, one reads of that the fusion ring of the
degenerate representations is
isomorphic to the tensor ring of $SU(2)\times SU(2)$. For $c=1$ one can
put $p=q=s=1$. The fusion
ring of the degenerate representations reduces to the fusion ring of
$SU(2)$.

Now let us consider the fusion of generic representations with ground
states $v_1,v_2$ and lowest
weight eigenvalues $h_1,h_2$ of $L_0$. Here, $V_1^s$ is given by the
action of the polynomials of
$L_1$ on $v_1$, without any relations, whereas $V_2^l$ is one
dimensional. Thus the lowest weight
space $V_1^s\otimes V_2^l$ is isomorphic  to the ring of polynomials
$p(x)$ in one variable. When one
puts $x=h_1+h_2+z_1L_1$, the action of $L_0$ is just given by
multiplication with $x$. This
representation has infinite dimension, only its finite dimensional
factor spaces induce
representations in $\cal V$. Those are of course obtained by factoring
out the ideal generated by
some polynomial $f(x)$. The zeros of $f(x)$ yield the lowest weight
eigenvalues of $L_0$. If no
multiple zeros occur, one obtains a completely reducible representation.
Since the zeros can be chosen
arbitrarily, all possible values of the lowest weights are permitted in
the fusion product, but only
with multiplicity one. Multiple zeros yield reducible but indecomposable
representations, whose
$n$-point functions involve logarithms.

\bigskip
{\bf Open questions}
\smallskip
Here is a list of the major questions posed above.

Are there lowest weight representations of some $W$-algebra which have
finite quantum dimension but
are not quasirational?

Are there conformal quantum field theories in two dimensions whose
$W$-algebras are not quasirational?

Of similar difficulty may be the construction of universal algorithms
for the following problems:
Check if $W$-algebras or particular representations of some $W$-algebra
are quasirational. Find the
spurious representations of the zero mode algebra in the fusion product
algorithm described above.

An easier problem should be the proof that standard construction
procedures for $W$-algebras like
orbifolding or GKO reductions always yield quasirational $W$-algebras
when applied to such algebras.
\smallskip
{\bf Acknowledgement}
I thank W. Eholzer and R. H\"ubel for useful discussions and
suggestions.
\vfil\eject
{\bf References}
\bigskip
\smallskip
[FF 88] B.L.Feigin and D.B.Fuchs, J.Geom.Phys. 5, 209,1988
\smallskip
[FFK 89] G.Felder, J.Fr\"ohlich, G.Keller, CMP 124, 647, 1989
\smallskip
[FFK 90] G.Felder, J.Fr\"ohlich, G.Keller, CMP 130, 1, 1990
\smallskip
[FRS 89] K.Fredenhagen, K.H.Rehren, B.Schroer, CMP 125, 201, 1989
\smallskip
[G 93] M.Gaberdiel, Fusion in Conformal field theory as the
tensor product of the symmetry
algebra, DAMTP preprint, 1993
\smallskip
[KL 93] D.Kazhdan and G.Lusztig: Tensor structures arising from affine
Lie Algebras I,II, preprints,
1993
\smallskip
[MMS 88] S.D.Mathur, S.Mukhi, A.Sen, Phys.Lett. B213, 303, 1988
\smallskip
[V 88] E.Verlinde, Nucl.Phys. B300, 360, 1988
 \medskip

\vfill \eject

\end